\documentclass[12pt,preprint]{aastex}
\usepackage{graphicx,amsmath,amssymb}
\usepackage{natbib}

\begin{document}

   \title{Delayed soft X-ray emission lines in the afterglow of GRB\,030227}

%   \shorttitle{X-ray line emission in \emph{INTEGRAL}'s GRB\,030227}

   \author{D.~Watson,\altaffilmark{1} J.~N.~Reeves,\altaffilmark{2,3} J.~Hjorth,\altaffilmark{1} P.~Jakobsson\altaffilmark{1} and K.~Pedersen\altaffilmark{1}}
   \altaffiltext{1}{Astronomical Observatory, NBIfAFG, University of Copenhagen, Juliane-Maries Vej 30, DK-2100 Copenhagen \O, Denmark}
   \email{darach@astro.ku.dk}
   \altaffiltext{2}{Laboratory for High Energy Astrophysics, Code 662, NASA Goddard Space Flight Center, Greenbelt, MD 20771, USA}
   \altaffiltext{3}{Universities Space Research Association}

   \begin{abstract}
     Strong, delayed X-ray line emission is detected in the afterglow of
     GRB\,030227, appearing near the end of the \emph{XMM-Newton}
     observation, nearly twenty hours after the burst.  The observed flux in
     the lines, not simply the equivalent width, sharply increases from an
     undetectable level ($<1.7\times10^{-14}$\,erg\,cm$^{-2}$\,s$^{-1}$,
     $3\sigma$) to
     $4.1^{+0.9}_{-1.0}\times10^{-14}$\,erg\,cm$^{-2}$\,s$^{-1}$ in the
     final 9.7\,ks.  The line emission alone has nearly twice as many
     detected photons as any previous detection of X-ray lines.  The lines
     correspond well to hydrogen and/or helium-like emission from Mg, Si, S,
     Ar and Ca at a redshift $z=1.39_{-0.06}^{+0.03}$. There is no evidence
     for Fe, Co or Ni---the ultimate iron abundance must be less than a
     tenth that of the lighter metals. If the supernova and GRB events are
     nearly simultaneous there must be continuing, sporadic power output
     after the GRB of a luminosity $\gtrsim5\times10^{46}$\,erg\,s$^{-1}$,
     exceeding all but the most powerful quasars.
   \end{abstract}
   \keywords{ Gamma rays: bursts -- supernovae: general 
                -- X-rays: general
             }

   \maketitle

%
%--------INTRODUCTION---------
%
\section{Introduction\label{introduction}}

Analysis of the afterglows of long-duration $\gamma$-ray bursts (GRBs) have
finally shown their progenitors to be massive stars
\citep{2002Natur.416..512R,1998ApJ...494L..45P,1998Natur.395..670G}, with a
catastrophic endpoint that seems to produce both a GRB and a supernova
\citep{1999Natur.401..453B,hjorth2003,astro-ph/0304173}. Rapid follow-up
observations of GRBs at X-ray wavelengths have provided spectra of the
afterglows showing very high luminosity line emission.  Initially single,
emission lines believed to be Fe
\citep{2000ApJ...545L..39A,1998A&A...331L..41P,2000Sci...290..955P,2001ApJ...557L..27Y}
or Ni \citep{2002A&A...393L...1W} were reported and more recently transient
multiple emission lines from highly-ionised Si, S, Ar, Ca and possibly Mg
and Ni \citep{2002Natur.416..512R}. A careful analysis of the
\emph{Chandra} HETG grating spectra of the afterglow of GRB\,020813 has
indicated the presence of highly-ionized states of similar low-Z metals, in
particular S and Si at much lower equivalent widths than those observed in
GRB\,011211 \citep{astro-ph/0303539}.  Detection of soft X-ray lines is
critical since they require such high energies to produce them while at the
same time allowing fairly unhindered access to the GRB and its remnant with
an accuracy unachievable in hard X- and $\gamma$-rays.

%
%--------OBSERVATIONS---------
%
\section{Observations and data reduction\label{observations}}

\emph{XMM-Newton} \citep{2001A&A...365L...1J} began observing the error-box of GRB\,030227 eight hours
after the burst (for thirteen hours) and for the first time a GRB detected
by \emph{INTEGRAL} \citep{2003SPIE.4851.1104P} was localized to within a few
arcseconds \citep{2003GCN..1901....1L}.
Three exposures were made with the EPIC cameras, the first two interrupted
by high background events. The effective time of the first exposure was less
than 1\,ks for each EPIC camera and is contaminated by high particle
background, it was therefore not considered for spectral analysis. The
second exposure spanned 3.6\,ks for the EPIC-MOS cameras and 7.2\,ks for the
EPIC-pn and also suffered from a relatively high background rate. The third
exposure spanned 32.4\,ks for the MOS cameras and 30.9\,ks for the pn. The
background rate for this exposure decreased to a low level after a few
thousand seconds.  Data from the MOS and pn cameras are consistent allowing
for cross-calibration uncertainties of <~15\% between the instruments.
Because of the extra free parameters introduced by allowing for systematic
differences between instruments and the much greater sensitivity of the
EPIC-pn detector the pn data were used for spectral fitting. The final fit
results were then checked against the MOS data and found to be consistent.

The data were divided into four time segments to examine spectral evolution.
The first segment corresponded to the second exposure, the remaining three
comprising consecutive 10\,ks, 10\,ks and 10.9\,ks parts of the third
exposure, giving effective exposure times of 5.7\,ks, 8.7\,ks, 9.0\,ks and
9.7\,ks respectively.  The data reduction followed a standard procedure
similar to that outlined in
\citet{2002A&A...395L..41W} except that the data were processed and reduced
with the XMM-Newton Science Analysis Software version 5.4.1.  A spectral
binning using a minimum of 20 counts per bin was used.  Consistent results
were obtained using minima of 10, 12, 20 and 25 counts per bin, as expected
\citep{1998ApJ...500..893Y} and using background spectra from different
regions on the detector.

%
%--------RESULTS--------------
%

\section{Results\label{results}}

This is the first afterglow discovered for an INTEGRAL GRB and had an
average 0.2--10.0\,keV flux of
$8.7^{+0.7}_{-1.3}\times10^{-13}$\,erg\,cm$^{-2}$\,s$^{-1}$, decaying with a
power-law slope of $-1.0\pm0.1$ with no strong evidence for deviations from
this decay rate \citep{2003ApJ...590L..73M}.

As outlined above, the \emph{XMM-Newton} data were divided into four time
segments, each about 10\,ks long, to examine spectral evolution. The
complete spectrum of the afterglow, and each time segment individually, can
be fit with a power-law and require absorption well in excess of the
Galactic value, with absorption consistently around twice the Galactic
foreground column density \citep[the Galactic hydrogen absorbing column in
this direction is $22\times10^{20}$\,cm$^{-2}$, though the uncertainty on
this value may be as large as 50\%,][]{1990ARA&A..28..215D}.

\subsection{Spectral emission features}
\begin{figure}
 \includegraphics[angle=-90,width=\columnwidth,clip=]{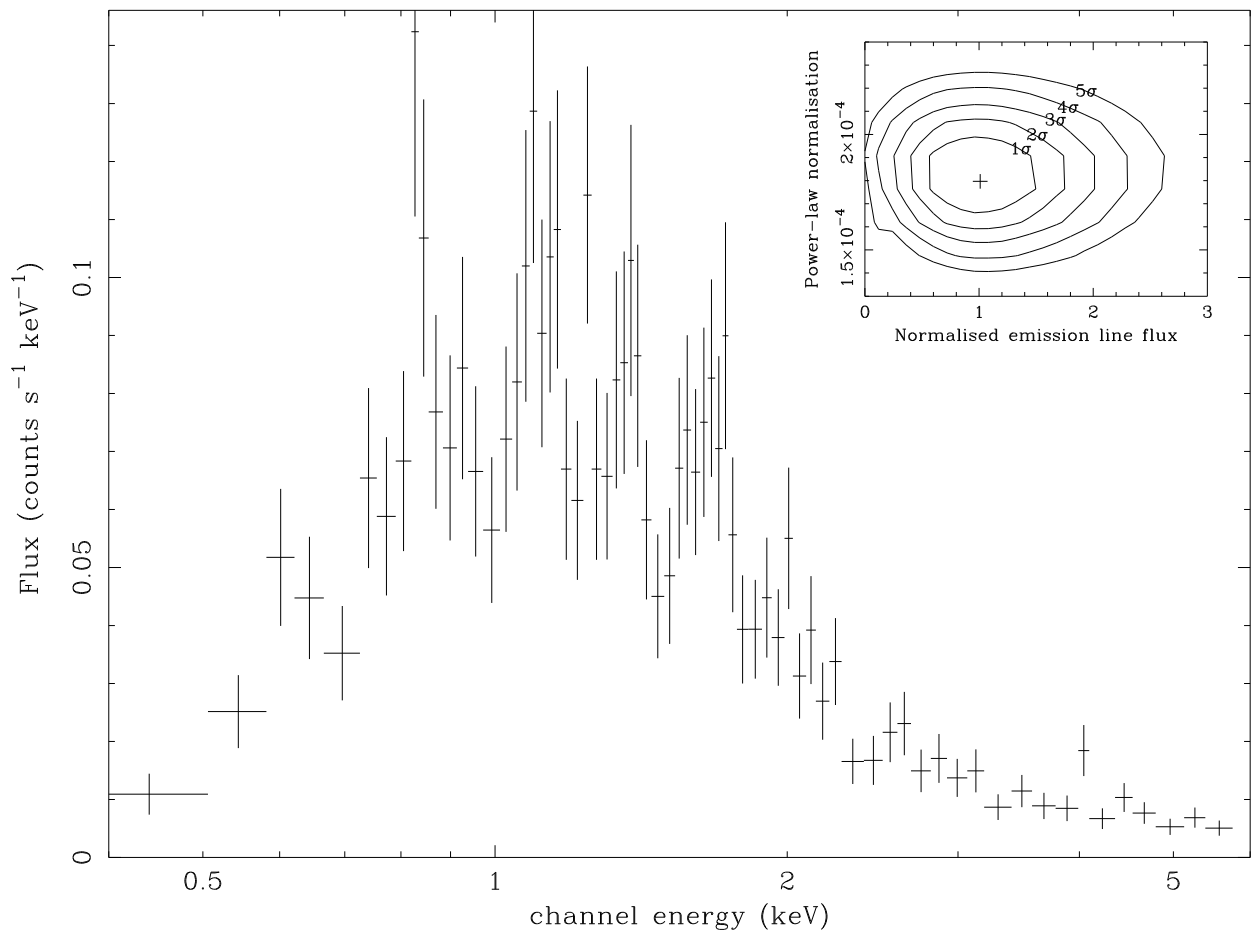}
 \caption{EPIC-pn spectrum of the final 9.7\,ks exposure of the afterglow
          of GRB\,030227.  The effects of the detector response have not
          been unfolded out of the data, however the emission lines are
          still clearly visible between 0.6 and 2\,keV.  The data are also
          not fit in this figure in order not to lead the eye.
          \emph{Inset} Statistical confidence contours for the normalised
          emission line flux against power-law normalisation for five
          parameters of interest.
         }
 \label{prob_contours}
 \label{untouched_spectrum}
\end{figure}

\begin{figure}
  \includegraphics[angle=-90,width=\columnwidth,clip=]{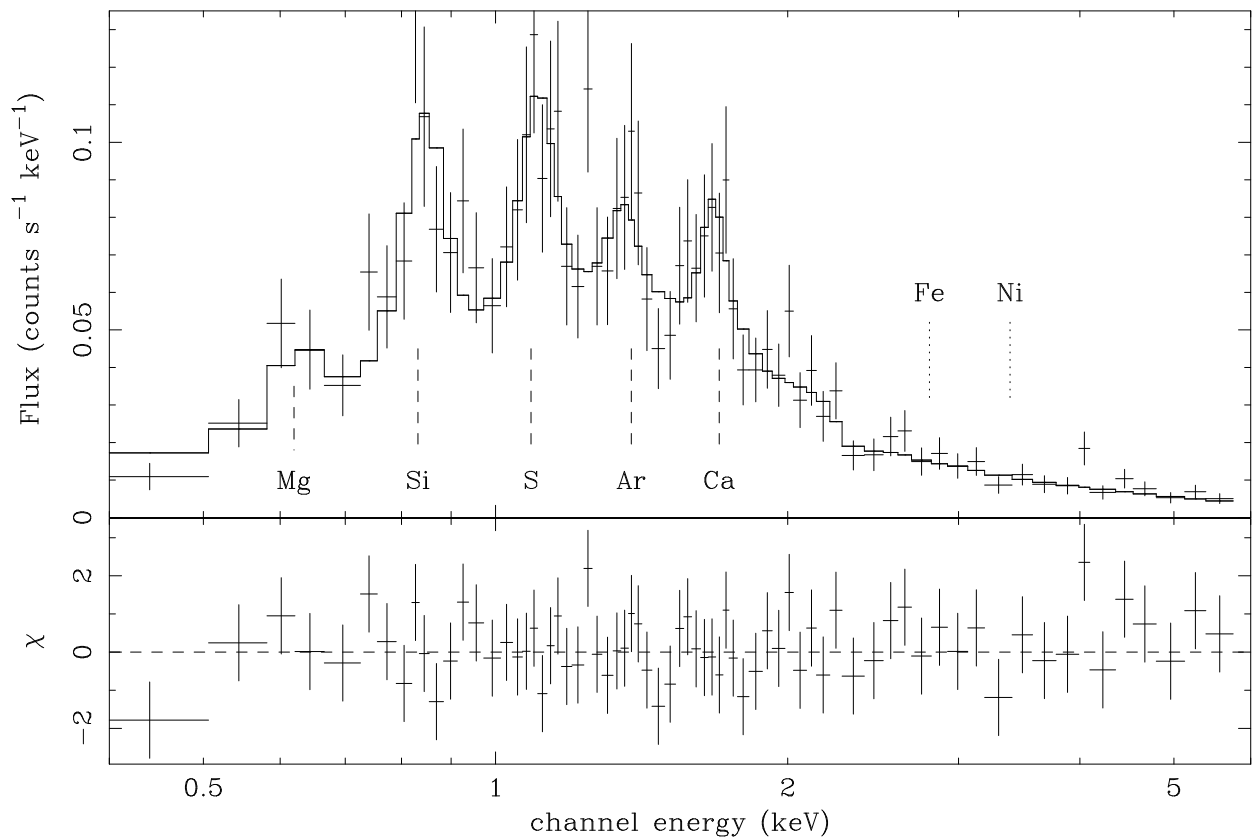}
  \caption{
           EPIC-pn spectrum of the last 11\,ks of the afterglow observation. 
           The data are fit with an absorbed power-law and five Gaussian
           emission lines. The K$\alpha$ lines of hydrogenic Mg, Si, S, Ar
           and Ca, redshifted by $z=1.39$ are marked with dashed lines. The
           nominal energies where the K$\alpha$ lines of hydrogenic Fe and
           Ni would be expected (also redshifted by $z=1.39$) are marked
           with dotted lines.  The continuum is smooth at these energies; no
           Fe, Co or Ni emission is detected.
          }
  \label{nomodel_spectrum}
% \end{minipage}
\end{figure}

However the spectrum evolves during the observation, showing emission lines
(see Fig.~\ref{untouched_spectrum}) at observed energies of
$0.62^{+0.03}_{-0.02}$, $0.86^{+0.02}_{-0.03}$, $1.11\pm0.02$,
$1.35^{+0.04}_{-0.03}$ and $1.67\pm0.04$\,keV only in the final $\sim10$\,ks,
approximately 70\,ks after the GRB.
The absorbed power-law fit ($\chi^2$/d.o.f. = 85.2/71) is clearly improved
in the last segment by the addition of the line emission
(Fig.~\ref{nomodel_spectrum}).
Adding five narrow Gaussian emission lines
(with energy and flux as free parameters) to a power-law with variable
slope, normalization and absorption, gave an improvement in the fit of
$\Delta\chi^2=33.3$, corresponding to a null hypothesis probability of
0.04\%.  Table~\ref{tab:lines} contains the individual best-fit line
parameters and significances.

Since four spectra were examined, a very
conservative estimate of the significance of these line features is
therefore given by the $f$-test probability for the addition of five
Gaussian lines to an absorbed power-law over four independent trials, which
is 0.15\% ($3.1\sigma$).  A more realistic estimate of the significance is
given by a fit to an \emph{a priori} expected model, i.e. that used by
\citet{2002Natur.416..512R} to characterise the line emission in
GRB\,011211.  In this model one expects to observe the Hydrogen-like
emission lines of Mg, Si, and S and the Helium-like lines of Ar and Ca at an
arbitrary redshift.  The trial-corrected $f$-test null hypothesis
probability from this model is $6\times10^{-5}$ ($4.0\sigma$).  Finally, one
can simply search the parameter space directly for the error on the total
line flux, for the relevant parameters of interest (power-law normalisation
and slope, absorption, redshift and the total emission line flux).  This
yields a probability of $3\times10^{-7}$ or $4.7\sigma$ ($4.4\sigma$ for
four trials) for zero emission line flux (see inset in
Fig.~\ref{prob_contours}).

\begin{deluxetable}{llccccc}
 \tablecaption{Best-fit quantities for the line emission.\label{tab:lines}}
 \tablecomments{Column~1 lists the probable line identifications.  For Ar and Ca, a
               helium-like ionization state seems more probable, giving a better match to
               the inferred redshifts (column~3) for the lighter elements.  Columns~2, 4 and 5 are the
               observed line energies, unabsorbed fluxes and observed equivalent widths respectively,
               while the fifth and sixth columns give the statistical significance
               (percentage probability and equivalent Gaussian standard deviations) of each
               line detection (based on the $f$-test comparison between the best-fit
               model with and without the line).
              }
 \tablehead{\colhead{Line} & \multicolumn{1}{c}{Energy} & \colhead{Line $z$} & \colhead{Unabs.\ Flux}               & \colhead{EW (eV)}    & \multicolumn{2}{c}{Sig.}\\
            \colhead{ID}   & \multicolumn{1}{c}{(keV)}  & \multicolumn{3}{c}{($10^{-14}$\,erg\,cm$^{-2}$\,s$^{-1}$)\phn\phn}                        & \colhead{\%} & \colhead{$\sigma$}}
 \startdata
  \ion{Mg}{12}   & $0.62^{+0.03}_{-0.02}$ & $1.35$ & $9.1^{+7.9}_{-6.3}$ & $211^{+182}_{-146}$ & 97     & 2.2\\[4pt]
  \ion{Si}{14}   & $0.86^{+0.02}_{-0.03}$ & $1.32$ & $4.1^{+2.7}_{-1.8}$ & $128^{+83}_{-57}$   & 99.98  & 3.8\\[4pt]
  \ion{S}{16}    & $1.11^{+0.02}_{-0.02}$ & $1.34$ & $2.4^{+0.8}_{-1.0}$ & $93^{+33}_{-39}$    & 99.96  & 3.5\\[4pt]
  \ion{Ar}{18}   & $1.35^{+0.04}_{-0.03}$ & $1.44$ & $0.9^{+0.9}_{-0.6}$ & $43^{+42}_{-27}$    & 92     & 1.7\\
  (\ion{Ar}{18}) & & (1.31) & & & & \\[4pt]
  \ion{Ca}{20}   & $1.66^{+0.04}_{-0.04}$ & $1.45$ & $1.3^{+0.6}_{-0.8}$ & $76^{+36}_{-44}$    & 99     & 2.5\\
  (\ion{Ca}{19}) & & (1.34) & & & & \\
%  \ion{Fe}{26}   & $2.6^{+0.3}_{-0.2}$    & $1.58^{+0.11}_{-0.10}$ &                    &                     &       & \\
 \enddata
\end{deluxetable}

\subsection{Comparison with previous X-ray emission lines}
The detection of soft X-ray line emission in GRB\,011211 has been
criticised on the basis of possible systematic errors
\citep{2003ApJ...583L..57B} and the level of statistical significance
\citep{2003MNRAS.339..600R}. Two later reanalyses of the data have been
unable to reproduce the systematic problems at all
\citep{2003A&A...403..463R,2003MNRAS.339..600R} and they appear to be due to
non-standard event selection by \citet{2003ApJ...583L..57B}.
%\footnote{Interestingly, we have been
%able to reproduce the spurious background line found by
%\protect\citet{2003ApJ...583L..57B} by including bad events, excluded in
%standard processing, lying on a \emph{vertical} chip edge physically
%unrelated to the location of the GRB afterglow (which lay near the
%horizontal chip edge) or the background regions chosen by
%\protect\citet{2002Natur.416..512R}.}
Concerns regarding the statistical
significance have been addressed by \citet{2003A&A...403..463R}.

Similar critical analysis has yet to be addressed to the other claims of
X-ray line detections.  An interesting comparison of the line detections
made to date is the number of photons detected only in the line emission. 
For instance in GRB\,991216, $\sim25$ photons \citep{2000Sci...290..955P};
GRB\,000214, $\sim35$ photons \citep{2000ApJ...545L..39A}; GRB\,020813,
$\sim60$ photons \citep{astro-ph/0303539}; GRB\,011211, $\sim115$ photons
\citep{2003A&A...403..463R}. In all of these bursts, the number of counts
detected per individual emission line has been $\sim25$--40.  In these
observations we detect $\sim210$ line photons, nearly twice as many as in
GRB\,011211 and nearly an order of magnitude more than GRB\,991216; the
number of counts for the brighter individual emission lines here is
$\sim60$.

%in units of ph/cm^2/s, GRBs
%991216 (3.2e-5, 9700s, ACIS-S orders -1, +1, HE + ME @ 3.5keV)
%        36 (HEG), 48 (MEG)
%        26 counts
%000214 (9e-6, 50000s, LECS + MECS @ 4.7keV)
%        75 (2+3 unit MECS, depends on off-axis position)
%        34 counts
%020813 (4.6e-6 + 6.5e-6 =~ 1.1e-5, 76800)
%     @  1.31,           1.01
%        30 (HEG) 67 (MEG), 11(HEG) 41 (MEG)
%        34              26  = 60
%011211 (1.0 + 1.0 + 0.8 + 0.4 + 0.2 = 3.35e-5, 5000s, EPIC-pn
%     @  0.44, 0.71, 0.88, 1.22, 1.46 keV )
%        290,  737,  884,  1003, 1044 cm^2
%        15  + 37  + 35  + 20  + 10 = 117 photons
%030227 (0.4 + 0.8 + 0.6 + 0.24 + 0.37 = 2.4e-5 for 9725s, EPIC-pn
%     @  0.62, 0.86, 1.11, 1.35,  1.66 keV )
%        612,  869,  978,  1022,  1019 cm^2
%        24  + 68  + 57  + 24   + 37 = 210 photons

%
%--------DISCUSSION-----------
%
\section{Properties of the line emission\label{discussion}}

The lines are remarkably similar to those observed in GRB\,011211
\citep{2002Natur.416..512R}---they correspond well to the Hydrogen- and
Helium-like lines of Mg, Si, S, Ar and Ca.  The median redshift of these
lines is $z=1.35$, where the ions are assumed to be Hydrogen-like; however
the ionisation states in the heavier elements, Ar and Ca, may be dominated
by the He-like ions (see Table~\ref{tab:lines}), resulting in a slighter
lower median redshift ($z=1.34$). The redshift determination is very robust;
the relative line-spacing is only correct for lines due to Mg, Si, S, Ar and
Ca and excludes the possibility of degeneracy in the redshift solution. An
outflow velocity close to $\sim0.1c$ is expected for the X-ray plasma
\citep{2002Natur.416..512R,astro-ph/0303539} which is similar to that
observed in optical GRB outflows \citep{hjorth2003}. The GRB progenitor
redshift is therefore expected to be $z\simeq1.6$.

The line emission alone accounts for about 6\% of the total flux in the
final ~10\,ks, which is an unabsorbed flux of
$1.8\times10^{-13}$\,erg\,cm$^{-2}$\,s$^{-1}$.  Assuming a redshift of
$z=1.6$ and a flat cosmology where H$_0=75$\,km\,s$^{-1}$\,Mpc$^{-1}$ and
$\Omega_\Lambda=0.7$, this translates to an isotropic line luminosity of
$2.6\times10^{45}$\,erg\,s$^{-1}$ or a total energy of at least
$1\times10^{49}$\,erg, implying that the lower limit to the energy required
to produce the lines is
$\sim2\times10^{50}$\,erg \citep{2002A&A...389L..33G}, within a factor of two of the total
$\gamma$-ray energy in the burst \citep{2001ApJ...562L..55F}.

\subsection{Iron, Cobalt and Nickel}
As with GRB\,011211 there is no evidence for emission from iron; the
$3\sigma$ upper-limit to the observed equivalent width for hydrogenic Fe is
175\,eV.  However, in this afterglow there is no evidence for Ni or Co
emission either (equivalent width $<140$\,eV).  In order to characterise the
relative abundances, a pure (absorbed) collisionally-ionised plasma model
was fit (without a power-law component).\footnote{Adding a power-law to the
thermal fit does not change the plasma temperature significantly, but makes
the absolute abundances harder to constrain by adding two extra free
parameters; however, the $2\sigma$ Fe abundance upper-limit is still an
order of magnitude less than the light metal abundance even after adding an
underlying power-law to the collisionally-ionised plasma model.} The
best-fit redshift for this model was $z=1.39^{+0.03}_{-0.06}$.  The minimum
($3\sigma$) light metal abundance was 24 times the solar abundance, compared
to upper limits of only 1.6 and 18 for Fe and Ni respectively.  The
implication of these upper limits is that there is only enough Fe, $^{56}$Ni
and $^{56}$Co to produce an order of magnitude lower abundance of iron than
that found for the lighter metals.  This is important since $^{56}$Ni is
produced in large quantities in supernovae and decays to $^{56}$Co and then
$^{56}$Fe and none of these products is observed. A low abundance of Fe, Co
or Ni is not anticipated in the standard models of GRB progenitors
\citep{2003ApJ...586.1254P,1998ApJ...507L..45V}, suggesting a bias towards
observing emission from the outer layers of the progenitor star and that
very little Ni has been dredged up to these layers by turbulent mixing by
the time the line emission is excited.

It has been suggested \citep{2002ApJ...572L..57L} that a high Fe abundance
may still be present while the Fe K$\alpha$ emission is suppressed by Auger
auto-ionisation.  A model based on reflection from an optically-thick medium
was examined.  Available models \citep[e.g.][]{2001ApJ...559L..83B} are
dominated by Fe emission, do not include some abundant metals (S, Ar, Ca)
and cannot give a reasonable fit to the data at low energies; however the
model and data were compared at high energies in the relevant ionisation
regime.  One can indeed strongly depress the Fe K$\alpha$ lines, however,
the model cannot fit the continuum at energies above 7\,keV (in the rest
frame) due to large Fe absorption. It appears unlikely that this mechanism
can both suppress the Fe or Ni emission and produce a smooth continuum at
high energies.  It is therefore concluded that Si and other $\alpha$-burning
products but not Fe, Co or Ni are highly abundant in the line-emitting
plasma.

\subsection{Photo-ionisation}
The line emission may result from photo-ionisation rather than collisional
ionisation.  In this case the plasma may have quite different properties.  A
model for a photo-ionised plasma was developed with XSTAR \citep{xstar}
allowing variable abundances of the most abundant elements and was added to
the absorbed power-law model and fit to the data from the final segment of
the observation. The data were as well-fit by this model ($\chi^2$/d.o.f. =
59.3/62) as by the collisionally-ionised model ($\chi^2$/d.o.f. = 61.3/67),
though requiring more free parameters.  The model requires a moderately high
ionisation parameter, $\log\Xi\sim3$.  The best-fit redshift
($1.38^{+0.05}_{-0.03}$) and elemental abundance ratios (high light metal
abundances, low Fe) are similar to those derived from the
collisionally-ionised model above, confirming that these results are
independent of the assumed emission model.

\begin{figure}
 \includegraphics[width=\columnwidth,clip=]{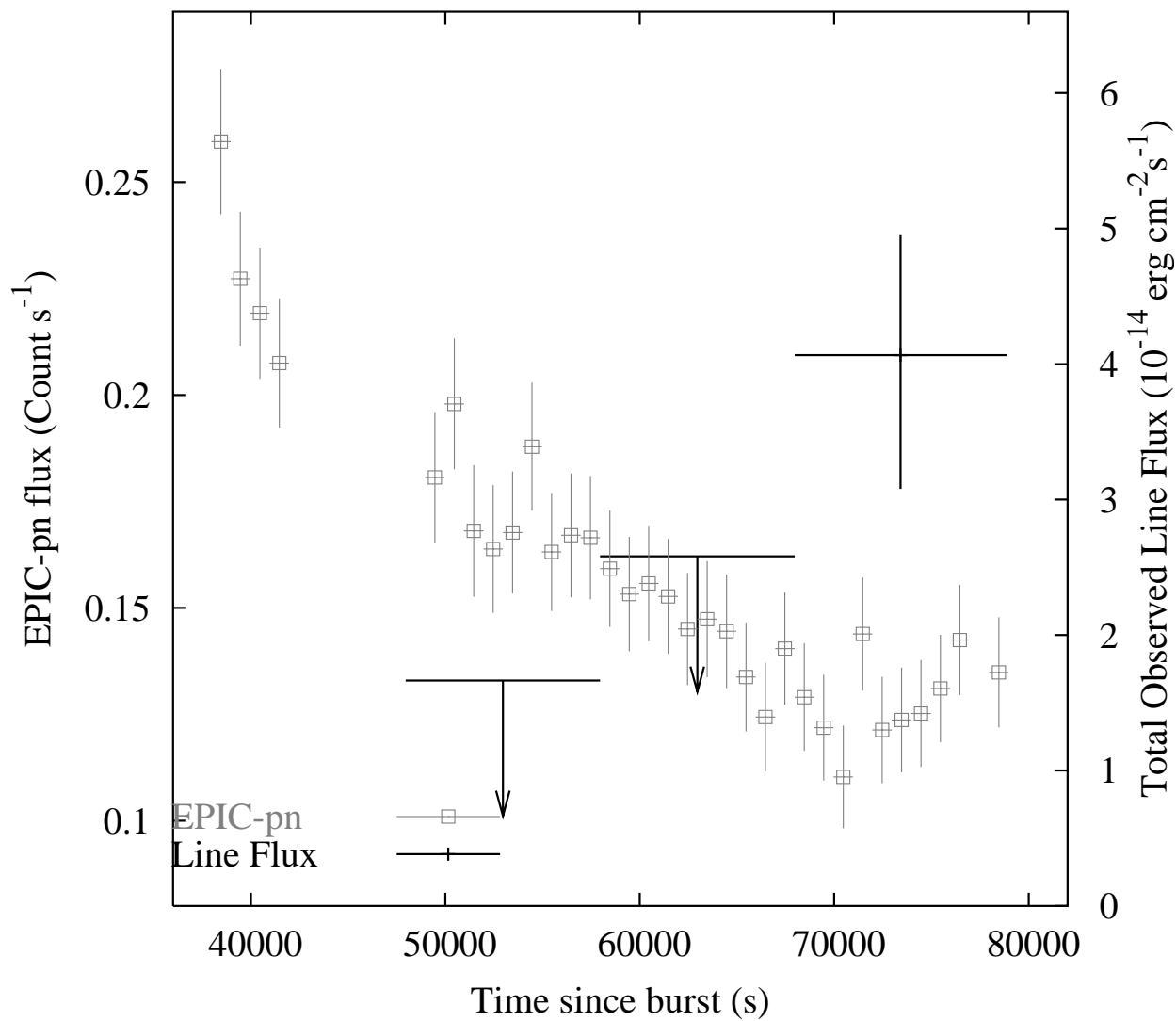}
 \caption{
          EPIC-pn 0.2--10.0\,keV lightcurve for the afterglow of GRB030227
          (grey, left axis) and the variation of the total line flux (with
          $3\sigma$ upper limits), for the longest EPIC-pn exposure (black,
          right axis).  The intrinsic flux in the lines increases from an
          undetectable level and becomes noticeable about 70\,ks after the
          GRB, corresponding to $\sim27$\,ks in the progenitor rest frame.
         }
 \label{line_variability}
 \label{lightcurve}
\end{figure}

If the underlying power-law slope does not change significantly during
the observation, it is possible to discount the sudden appearance of thermal
emission in the last 11\,ks of the observation, since the thermal flux would
have to be about half the total flux (in the case where the spectral index
and absorption are fixed at the level required to fit the full
dataset) and the sudden appearance of emission of this magnitude is ruled
out by the X-ray lightcurve (Fig.~\ref{lightcurve}).
The extra emission component must therefore either appear gradually or be
non-thermal in origin, strengthening the case for photo-ionisation.

\section{Implications of the delay}
The line emission in this afterglow is detected only in the final segment of
the observation (Fig.~\ref{line_variability}), implying that the lines not
only fade, as observed in GRB\,011211, but also appear a significant time
after the GRB, in this case six or seven hours in the rest frame.  There are
two alternative explanations for delayed line emission and each specifies
both the geometry and the delay in the central power output.

The first alternative involves direct excitation of the lines by the GRB
event; they must be delayed by taking a much longer path to the observer
than the $\gamma$-rays (reverberation).  In the second case there must be
continuing, sporadic injection of energy after the GRB has finished.

In a reverberation-dominated scenario the minimum distance from the GRB to
the line-emitting plasma can be estimated.  It is $9\times10^{14}$\,cm
(where the off-axis angle is $90\degr$).  If the plasma is flowing out from
the stellar remnant at $\sim0.1c$ \citep{2002Natur.416..512R,hjorth2003,astro-ph/0303539} it implies a delay of at least 3.5
days between the event causing the matter outflow (presumably a supernova)
and the GRB.  If the line emission is produced by the GRB jet itself, the
line-emitting plasma must be very near or in the edge of the jet.  The
off-axis angle must then be much less than $90\degr$ implying a longer time
delay between the matter outflow and the GRB.  The jet opening angle (and
hence the off-axis angle for the material) can be constrained.  Using the
method of \citet{2001AJ....121.2879B} the equivalent isotropic energy was derived for
this burst (extrapolating the INTEGRAL-SPI spectrum) and the actual
$\gamma$-ray energy release was assumed to be $5\times10^{50}$\,erg \citep{2001ApJ...562L..55F}. The ratio of these numbers implies a half-opening angle of
$\sim15\degr$ for the jet.  At this angle, the inferred distance is
$2\times10^{16}$\,cm implying a delay between the matter expulsion and the
GRB of about 80 days.  A supernova considerably prior to the GRB (e.g.\ a
`supranova') would be a natural interpretation in this scenario
\citep{2003A&A...399..913L}. It should be noted that for delay due to
reverberation as outlined above, the line-emitting material would have to be
enriched in light elements, poor in iron, nickel and cobalt and concentrated
in dense clumps slightly off-axis from the GRB jet but generally located
around this axis, otherwise the total ejected mass is prohibitively large.

Very recent observations of a nearby GRB (030329) show clear evidence of a
supernova in the optical spectra, constraining the time-delay between the
supernova and the GRB to be less than a few days
\citep{hjorth2003,astro-ph/0306155}. If the result is general and all long-duration
GRBs have the same progenitors, as appears likely, then the delay between
supernova and GRB required in the reverberation scenario described above
probably disqualifies it.  An analysis of a different reverberation scenario
has been made by \citet{2003ApJ...584..895K}, where $\gamma$- and hard
X-radiation from the GRB and the early afterglow is reflected back onto the
outer layers of the expanding supernova.  This concept has the advantage
that it does not require a delayed two-stage explosion sequence (SN, GRB) to
produce the reverberation.  Furthermore, it naturally explains the lack of
emission from heavier ions (Fe, Co, Ni) if the outer layers of the SN ejecta
are dominated by lighter metals.  The principle difficulty, as pointed out
by \citet{2003ApJ...584..895K}, is that it is hard to produce X-ray line
luminosities $>10^{48}$\,erg as directly observed here.

Finally we turn to continuing energy injection to explain the delayed X-ray
lines.  In the collapsar model \citep{1999ApJ...524..262M} where the radius
of the dense matter must be $\lesssim10^{13}$\,cm (the size of an exploding
Wolf-Rayet star) and a single event (the GRB) must produce all the observed
features, the reverberation time is too short to account for the observed
delay.  In order to produce emission lines it has been proposed
\citep{2000ApJ...545L..73R} that a strong, post-GRB source is reflected off
the internal edges of a cavity evacuated in the star by the GRB jet
\citep{2003ApJ...586..356Z}. The enormous power in the X-ray lines implies
that the continuum luminosity onto the cavity edge must, at the very least,
be
$5\times10^{46}$\,erg\,s$^{-1}$ for a duration of a few thousand seconds.  A
source with this luminosity would equal the X-ray afterglow continuum
observed here and be clearly detected; the continuum source must therefore
either be obscured or be intrinsically anisotropic.  The more likely
proposition is an anisotropic source as expected for a hot accretion disk
\citep{2003ApJ...586..356Z} or a young, rapidly accreting pulsar with a
strong magnetic field \citep{2000ApJ...545L..73R}, both proposed as
consequences of a GRB in a massive star.  In normal pulsars the emission
axis is offset from the rotation axis, as is required here.  Furthermore a
natural corollary of the geometry in the case of the pulsar, if the emission
can be restricted to a cone-shaped beam from the poles, is that the inner
regions of the cavity wall, where the Fe/Ni abundance can be expected to be
highest, may not be illuminated by the pulsar beam, resulting in a
light-metal-rich, but Fe/Ni-deficient X-ray reflection spectrum due to the
laminar separation of elements in an aged, massive star.

\acknowledgments
The authors thank Fred Jansen for the allocation of discretionary XMM-Newton
time, the Science Operations Centre team for performing the rapid follow-up
observation, L.~Hanlon, D.~Lazzati, B.~McBreen, and E.~Ramirez-Ruiz for
discussions, and acknowledge benefits from collaboration within the EU FP5
Research Training Network, `Gamma-Ray Bursts: An Enigma and a Tool'. This
work was also supported by the Danish Natural Science Research Council
(SNF).

\end{document}